\documentclass[twocolumn]{pasj00}

\begin{document}
\SetRunningHead{K. Motogi et al.}{New Distance and Revised Natures of G5.89-0.39}
\Received{2010 August 14}
\Accepted{2010 December 15}

\title{New Distance and Revised Natures of High Mass Star Formation in G5.89-0.39}

\author{Kazuhito \textsc{motogi},\altaffilmark{1} %
Kazuo \textsc{sorai},\altaffilmark{1,2} %
Asao \textsc{habe},\altaffilmark{1,2} %
Mareki \textsc{honma},\altaffilmark{3,4} %
Hideyuki \textsc{kobayashi},\altaffilmark{3,4,5} %
and Katsuhisa \textsc{sato}\altaffilmark{4}} %
\altaffiltext{1}{Department of Cosmosciences, Graduate School of Science, Hokkaido University, N10 W8, Sapporo, Hokkaido 060-0810}
\email{motogi@astro1.sci.hokudai.ac.jp}
\altaffiltext{2}{Department of Physics, Faculty of Science, Hokkaido University, N10 W8, Sapporo, Hokkaido 060-0810}
\altaffiltext{3}{Mizusawa VLBI Observatry, National Astronomical Observatory, 2-12 Hoshi-ga-oka, Mizusawa-ku, Oshu, Iwate 023-0861}
\altaffiltext{4}{Department of Astronomical Sciences, Graduate University for Advanced Studies, 2-21-1 Osawa, Mitaka, Tokyo 181-8588}
\altaffiltext{5}{Department of Astronomy, Graduate School of Science, The University of Tokyo, 7-3-1 Hongo, Bunkyo-ku, Tokyo 113-0033}


%


\maketitle

\begin{abstract}
We report on the astrometric observations of the 22 GHz H$_{2}$O masers 
in the high mass star-forming region G5.89-0.39 with VERA (VLBI Exploration of Radio Astrometry). 
Newly derived distance of 1.28$^{+0.09}_{-0.08}$ kpc is the most precise and significantly nearer than previous values. 
We revised physical parameters and reconsidered nature of G5.89-0.39 based on the new distance as follows. 
(1) The ionizing star of the ultra compact (UC) H$\emissiontype{II}$ region is a late O-type (O8 - 8.5) zero age main sequence (ZAMS) star, 
consistent with previously established limits based on its infrared spectral line emission. 
(2) Crescent-like maser alignment at the position of the O type ZAMS star may trace accretion disk (or its remnant), 
which suggests that the star is still young and before complete evaporation of circumstellar materials. 
(3) Although the revised mass for the east-west outflow has been reduced, it still quite large (100 $\MO$) which indicates that a significant
fraction of the mass is entrained material and that the dynamical age significantly underestimates the actual outflow age. 
Our newly-derived distance emphasizes that G5.89-0.39 is one of the nearest targets to investigate ongoing high-mass star formation 
and evolution in a compact cluster containing a young O-type star. 
\end{abstract}

\newpage

\section{Introduction}
A systematic and consistent scenario of high mass star formation has not been constructed yet. 
In spite of enormous and intense works, there are many hypotheses and unsolved issues (e.g., \cite{Zinnecker2007} and reference therein). 
But, several recent theoretical studies have suggested that high mass star formation can be achieved via mass accretion (e.g., \cite{Krumholz2009}). 
This hypothesis seems to be consistent with observational signatures of massive accretion 
disk and torus around high mass protostellar object (HMPO) or protocluster (e.g., \cite{Beuther2009}). 
Upcoming Atacama Large Millimeter / submillimeter Array (ALMA) will be able to resolve such a circumstellar structure enough 
and provide us more quantitative information about a specific accretion mechanism onto an individual HMPO. 

Precise distance of a source is essential to quantitative discussions. 
However, most of the high mass star-forming regions, which will be targets for ALMA, are even located on the inner Galactic plane, 
where source distances often contain significant uncertainty. 
Direct distance measurements by astrometric observations are quite important for such regions, 
in especial, highly accurate VLBI (Very Long Baseline Interferometry) astrometry is the only technique suitable for 
deeply embedded high mass star-forming regions where optical measurements are almost impossible. 
\citet{Hachisuka2006} has been first demonstrated great performance of VLBI astrometry using H$_{2}$O maser in W3(OH) region, 
and since then, several studies have achieved 10 micro-arcsecond ($\mu$as) accuracy for northern star-forming region (e.g., \cite{Moellenbrock2009}; \cite{Sato2010b}). 
In this paper, we report on an annual parallax measurement of H$_{2}$O masers in the high mass star-forming region G5.89-0.39 
with VERA (VLBI Exploration of Radio Astrometry; \cite{Kobayashi2008}). 

G5.89-0.39 (also known as W28A2) is one of the most famous, shell type ultra compact (UC) H$\emissiontype{II}$ region (e.g., \cite{Wood1989}). 
The O-type ionizing star has been detected as a near-infrared (NIR) point source by \citet{Feldt2003} inside the shell (hereafter Feldt's star). 
\citet{Acord1998} (hereafter ACW98) have directly measured dynamic angular expansion of the radio shell. 
Observed supersonic expansion and short dynamical age (600 yr) indicate that this small UCH$\emissiontype{II}$ region is just after the birth. 
G5.89-0.39 is also known to be a host of an extremely massive outflow which is centered on the shell (e.g., \cite{Acord1997}). 
The whole part of the shell is completely inside the outflow extent (e.g., \cite{Watson2007}). 
This also gives further support on a remarkable youth of the UCH$\emissiontype{II}$ region. 

Previously reported distances for G5.89-0.39 vary over a wide range (1.9 - 3.8 kpc; \cite{Hunter2008}, and reference therein). 
Almost all of them are measured through kinematic distance method, 
but at the Galactic longitude of 5$^{\circ}$.89, this method intrinsically contains large systematic error of kpc order. 
Although ACW98 tried to estimate the distance from the shell expansion without any Galactic rotation models, 
they still have adopted several assumptions for their modeling of the data. 
In this point of view, our direct distance measurement is very important to confirm the physical parameters of G5.89-0.39. 

\section{VERA Observations}
 VERA observations of the H$_{2}$O masers ($J_{K_{\mathrm{a}}K_{\mathrm{c}}}$ = 6$_{16}$--5$_{23}$) at 22.23508$\:$GHz 
 associated with G5.89-0.39 have been carried out at 9 epochs between 2007 November and 2009 May. 
 A summary of all observations is listed in table \ref{tb:Obs}, which contains observing dates, synthesized beam sizes in milli-arcsecond (mas), 
 beam position angles (PA) east of north, spectral resolutions, typical noise levels and dynamic ranges of synthesized images 
 and brief comments, if any, about system noise temperatures ($T_{\mathrm{sys}}$). 
 
 Each observation was made in VERA's dual-beam mode in which targeted maser source and phase calibrator (or position-reference source) 
 were observed simultaneously (\cite{Kawaguchi2000}; \cite{Honma2003}).
 The real-time, instrumental phase difference between the two beams was measured for a calibration 
 using the artificial noise sources during the observations \citep{Honma2008a}. 
 We chose J1755-2232 ($\alpha_{2000}$=\timeform{17h55m26.2848s}, $\delta_{2000}$=\timeform{-22D32'10.61656"}; \cite{Petrov2005}) 
 as a paired calibrator. This source is separated from G5.89-0.39 by 1$^{\circ}$.92 at a position angle of -140$^{\circ}$ east of north. 
 The flux density of J1755-2232 is about 150 mJy. No significant structure is seen in this point-like calibrator. 
 A bright calibrator, NRAO530 (=J1733-1304; Ma et al. 1998), was also scanned every 120 minutes as a delay and bandpass calibrator. 
 Each observation was made for about 6 hours, but total scan time for the targeted source pair was only about 2.5 hours
 because we observed not only G5.89-0.39--J1755-2232 pair but also another maser--calibrator pair alternately. 
 We will describe another source pair in forth coming paper. 
 
 Left-handed circular polarized signals were quantized at 2-bit sampling and filtered with the VERA digital filter unit \citep{Iguchi2005}, 
 after that, data were recorded onto magnetic tapes at a data rate of 1024 Mbps.
 There were 16 IF channels with 16 MHz band width where one IF was assigned to the maser lines 
 and other 15 IF of total 240 MHz were assigned to J1755-2232 and NRAO530. 
 The data correlation was performed with the Mitaka FX correlator \citep{Chikada1991}.
 Correlated data were divided into 512 and 64 spectral channels for the maser and calibrators, respectively.
 For the maser lines, the full 16 MHz data were used only in the fourth epoch, 
 and in other cases, we used 8 MHz which is covering whole maser emissions to achieve sufficient velocity resolution (0.21 km s$^{-1}$).
 
 The system noise temperatures were depended on weather conditions in each station and elevations.
 Typical $T_{\mathrm{sys}}$ value at a averaged elevation angle was varied 200 to 500 K for each station in a case of normal weather.
 Sometimes it exceeded 1000 K in a bad case for Ogasawara and Ishigaki station. 
 There was very high $T_{\mathrm{sys}}$ value ($\sim$ 5000 K) at Iriki station in the first epoch, 
 and hence, we performed phase-referenced imaging using other three stations in this epoch.
 Phase-referenced images of the targeted maser source were successfully obtained in 8 epochs 
 without the final epoch, where significant phase fluctuations were still remained after the fringe fitting for J1755-2232, 
 and then, all maser spots were completely defocused. 
 
\begin{table*}
 \centering
  \caption{Summary of the Obsevations}\label{tb:Obs}
   \begin{tabular}{ccccccc}\hline
   Date& Synthesized Beam & PA & Resolution & 1-$\sigma$ Noise$^{a}$ & DN$_{p}$$^{b}$ / DN$_{s}$ & Comments \rule[0mm]{0mm}{4mm}\\
       & (mas $\times$ mas) & ($^{\circ}$) & (km s$^{-1}$) & (Jy beam$^{-1}$) &($\sigma$) &for $T_{\mathrm{sys}}$\\ \hline
2007/11/5 & 1.92 $\times$ 0.71 & -24.33 &0.21& 0.82  & 9 / 32& $\sim$ 5000 K at IR$^{c}$\\
2008/1/12 & 1.88 $\times$ 1.02 & -16.23 &0.21 &0.35  & 11 / 166& $\sim$ 1000 K at IS\\
2008/3/14 & 2.15 $\times$ 0.88 & -20.15 &0.21 &0.22  & 15 / 402& --\\
2008/5/7 & 2.32 $\times$ 0.88 & -24.48 &0.42& 0.63  & 14 / 530&  500 - 2000 K at OG\\
2008/7/1 & 2.14 $\times$ 0.81 & -23.04 &0.21& 0.42  & 8 / 128& --\\
2008/11/11 & 2.62 $\times$ 0.84 & -26.16 &0.21& 0.29  & 11 / 144& 800 - 3000 K at OG\\
2009/2/6 & 2.02 $\times$ 0.92 & -16.82 &0.21& 0.17  & 15 / 255& --\\
2009/5/18 & 2.26 $\times$ 0.84 & -23.32 &0.21& 0.28  & 15 / 43& --\\
2009/9/13 & 2.21 $\times$ 0.78 & -19.95 &0.21& 0.25  & -- / 54&--\\ \hline
\multicolumn {7} {l} {$^a$ Typical value in self-calibrated images.}\\
\multicolumn {7} {l} {$^b$ The maximum dynamic ranges for phase-referenced (DN$_{p}$) and self-calibrated image (DN$_{s}$).}\\
\multicolumn {7} {l} {$^c$ IR, IS, OG:Iriki, Ishigaki, Ogasawara station, respectively.} \\
     \end{tabular}
\end{table*}
 
 \section{Data reduction}
 Data reduction was carried out using the NRAO Astronomical Imaging Processing System (AIPS) package. 
 Amplitude and bandpass calibrations were made for the targeted maser and J1755-2232 independently. 
 We calibrated clock parameters for each station using the residual delay of NRAO530.
 The tropospheric zenith delay offset was also calibrated by the modified delay-tracking data
 which were calculated based on the actual measurements of the atmospheric zenith delay 
 with the global positioning system (GPS) at each station \citep{Honma2008b}. 
 
 There were two different paths of the analysis depending on a purpose. 
 The one was phase-referencing (or 2--beam) analysis for a measurement of the annual parallax 
 and another was single-beam analysis which was suitable for searching all maser spots and studying internal motions of them.
 Here, the term 'maser spot' indicates a maser component seen in a single velocity channel, and by contrast, 
 the term 'maser feature' represents physical gas clump which is consist of several maser spots 
 which are detected in successive velocity channels and closely located each other. 
 We define it in the same way as \citet{Motogi2008}. 
 
 In the single beam analysis, we simply performed fringe fitting and self-calibration 
 using the brightest maser spot in the feature O1 (see section 4.3). 
 We then searched for all maser spots by a wide-field mapping under a 7-$\sigma$ detection limit.
 Total explored area was 5 $\times$ 5 arcsec$^{2}$ of each channel map centered on the ionizing star in G5.89-0.39.
 Annual aberration was also corrected for several masers which were significantly distant ($\gg$ 3$^{\prime\prime}$) 
 from the phase-referenced maser spot in this analysis,. 
 
 In the phase-referencing analysis, fringe fitting and self-calibration for J1755-2232 were done, and then, 
 obtained delay, rate and phase solutions were applied to the target visibilities 
 following the correction of measured instrumental phase difference between the two beams (see above). 
 The modified delay-tracking mentioned above was re-calculated with respect to each maser feature for accurate position measurement. 
 New delay-tracking center was always located within 10 mas from relevant maser feature. 

 The coherence in phase-referenced images were still significantly degraded at this stage. 
 Because of the low brightness and large separation angle of J1755-2232, and the low elevation angle during observations, 
 significant atmospheric zenith delay residuals, which is a main cause of the coherence loss, still remain.
 We, therefore, estimated and corrected this residuals with the image optimizing method described in Honma et al. (2007, 2008b).
 Estimated residuals were found to be within $\pm$3 cm in whole epochs ($\sim$ 1.5 cm on average).
 This is consistent with a typical case seen in VERA observations \citep{Honma2008b}.
 The dynamic ranges (or signal to noise ratios) of phase-referenced images showed 
 dramatic improvement with this correction. 
 But, even after this improvement, limited coherence and defocusing made 
 several faint spots undetectable.
 
 After these calibrations, synthesized image cube was finally made with all maser features in both analyses.
 Each cube had a field of view of 25.6$\times$25.6 mas$^{2}$ centered on relevant feature. 
 Imaging and deconvolution (CLEAN algorithm) was done in uniform weighting which provided the highest spatial resolution.
 The synthesized beam was about 2.2$\times$0.9 mas$^{2}$ with the position angle of -20$^{\circ}$.
 Typical image noise level was $\sim$ 400 mJy beam$^{-1}$ (1-$\sigma$) in the case of the single-beam analysis (see table \ref{tb:Obs}).
 It was significantly increased in phase-referenced case by a factor of 2 or 3, and up to 10 in the worst case.
 The absolute and relative positions of each maser spot were determined by an elliptical Gaussian fitting.
 The formal error in this fitting was typically 50 $\mu$as in RA and 100 $\mu$as in Dec. 
 This value is approximately equal to a uncertainty of relative positions between each maser spots.
 Overall discussion about the accuracy of absolute positions in each measurement is given in section 4.2.
 
 \section{Result}
 \subsection{Parallax Measurement}
 There were two strong maser features (feature O1 and C6 in section 4.3) successively detected for 8 epochs in phase-referenced maps. 
 We measured their absolute motions referenced on J1755-2232.
 These motions can be expressed by the sum of the annual parallax $\pi$ and proper motion of each feature $\mu$.
 The latter is usually assumed to be a linear and constant motion for simplicity (e.g., \cite{Nakagawa2008}). 
 This seems to be applicable for our case, 
 because the relative proper motion between two features are actually fitted by a linear motion (see section 4.3 and figure \ref{fig:inmotion}) and 
 the drifts of their line of sight velocities are well negligible ($<$ 0.2 km s$^{-1}$ ) during the two years. 
 
 A least square fitting was made with $\pi$ plus $\mu$ 
 to the right ascension offsets ($X \equiv  \Delta\alpha \times \mathrm{cos}\delta$ ) from the first epoch. 
 We performed our fittings in the same way as \citet{Hirota2008}.
 Here, whole detected maser spots in each feature (7 maser spots per feature) were distinctly used and an initial position of each maser spot, 
 $X_{0}$ and $Y_{0}$, was included as a fitting parameter. 
 Reduced $\chi^{2}$ was calculated as 
 \begin{eqnarray}
 \chi^{2} = \frac{1}{m}\sum^{}_{i} \omega_{i} (X_{i} - f(t_{i}))^{2}, \nonumber 
 \end{eqnarray}
 where $m$ and $\omega_{i}$ were the degree of freedom and a fitting weight for each data point, respectively. 
 We adopted a dynamic range of phase-referenced image (DN) as a fitting weight, 
 since it was a rough measure of the coherence loss caused by residual delay which was difficult to estimate directly.
 A weight for $i$--th data point was to be $\omega_{i}\propto$ $\mathrm{DN}_{i}$ 
 and scaled to make the total reduced $\chi^{2}$ to be $\sim$ 1. 
 If we assume the normal distribution, $1/\sqrt{\omega_{i}}$ is equal to the error variance of $i$--th data point. 
 
\begin{figure*}
  \begin{center}
    \FigureFile(155 mm,80mm){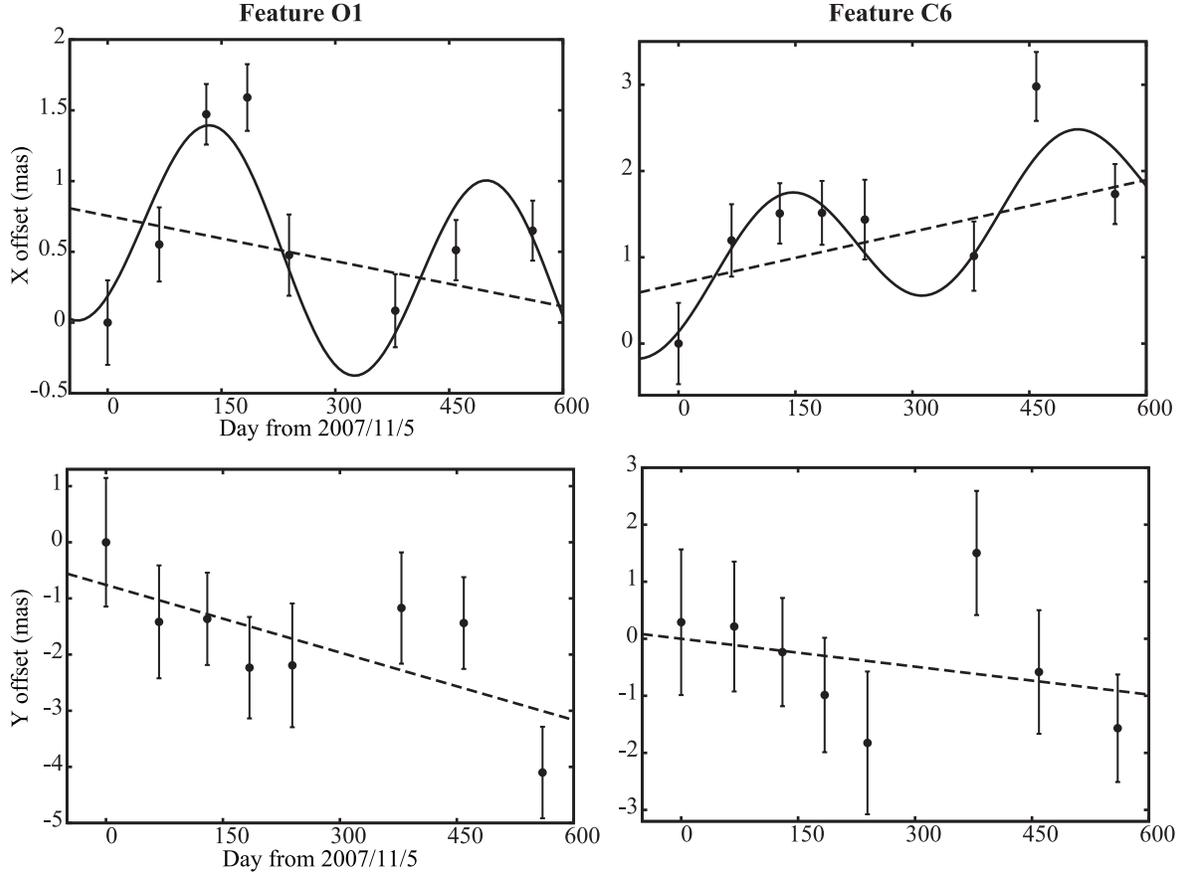}
  \end{center}
  \caption{The examples of parallax fitting for the brightest maser spot of feature O1 (left) and C6 (right). 
  Here, $X$ and $Y$ axis shows elapsed days from the first epoch and positional offsets 
  from $\alpha_{2000}$=\timeform{18h00m30.3066s}, $\delta_{2000}$=\timeform{-24D04'04.48649"}in mas, respectively. 
  The annual parallax (solid curve) and linear proper motion (dashed line) 
  were successfully fitted in $X$ direction (upper two panels). On the other hand, 
  only a linear proper motion was fitted in $Y$ direction because of the location of G5.89-0.39 (see text). 
  The errors in each panel are calculated from the fitting weights which have been scaled to make the total reduced $\chi^{2}$ to be $\sim$ 1.}\label{fig:Parallax}
\end{figure*}

 Figure \ref{fig:Parallax} shows the examples of the parallax measurement for G5.89-0.39.
 Note that only a linear proper motion was fitted to the declination offsets ($Y \equiv \Delta\delta$), 
 since G5.89-0.39 was located on the ecliptic plane ($\beta\sim\:$-0.6$^{\circ}$), 
 where the parallax in the declination degenerated into an order of 10 $\mu$as. 
 It is just comparable with the maximum accuracy of VERA's dual beam astrometry \citep{Honma2007} and hard to detect. 
 In each figure, the associated error bars indicate $1/\sqrt{\omega_{i}}$ values. 
 The $X$ were well fitted by a linear motion (dashed line) and parallax (solid curve), by contrast, 
 large dispersion from the best-fit linear motion is clearly seen in cases of $Y$.
 The error in $Y$ ($\sim$0.9 mas) is, in fact, much larger than that in $X$ ($\sim$ 0.4 mas) 
 as well as several other observations with VERA (e.g., Hirota et al. 2007). 
 Detailed analysis about these errors is given in next subsection. 
 
 Table \ref{tb:parallax} presents a summary of these fittings. 
 Presented $\sigma$ values are rms deviations of post-fit residuals which are good measures of typical error for each measurement. 
 Because both of the feature O1 and C6 seem to be associated with a single UCH$\emissiontype{II}$ region, 
 we can assume that they show the same annual parallax within the astrometric accuracy of VERA. 
 Thus, the best-fit annual parallax was estimated as $\pi$ = 0.78 $\pm$ 0.04 mas, corresponding a distance of 1.28$^{+0.07}_{-0.06}$ kpc from the sun, 
 by the weighted-means of those for the two features. 
 
 One can see that some data points do not included within 1-$\sigma$ error range in figure \ref{fig:Parallax}. 
 This can be due to inequable data quality in each observing epoch. 
 We have, hence, performed bootstrap analysis to test such an effect, where 
 data for each maser spot have been randomly resampled 10000 times and a least square fitting has been iteratively done for each dataset. 
 
 Obtained frequency distributions of $\pi$ have been well fitted by Gaussian. 
 Detailed parameters are presented in table \ref{tb:bootstrap}. 
 Resultant error of $\pm$ 0.05 mas is slightly larger than that of simple fittings but almost consistent with it, and 
 we finally take it as conclusive 1-$\sigma$ error of our parallax measurement. 
 We also emphasize that even if we adopt 3-$\sigma$ error ($>$ 99 $\%$ confidence), 
 equivalent distance range is still significantly nearer than previously reported values. 
 
 \begin{table*}
 \centering
  \caption{Summary of simple parallax fittings}\label{tb:parallax}
   \begin{tabular}{cccccccc}\hline
   Feature&V$_{\mathrm{LSR}}$ (km s$^{-1}$)&&$\pi$ (mas) &D (kpc) & $\sigma^a$ (mas) & $\mu^b$ (mas yr$^{-1}$) \rule[0mm]{0mm}{4mm}\\ \hline\hline
  O1&9.4&$X$& 0.79 $\pm$ 0.05 &1.27 $\pm$ 0.08& 0.27 & -0.43 $\pm$ 0.09 \rule[0mm]{0mm}{4mm}\\ 
  &&$Y$&-- & -- & 0.78 & -1.43 $\pm$ 0.29\\ 
  C6&11.0&$X$& 0.77 $\pm$ 0.09 &1.30$^{+0.17}_{-0.14}$ & 0.39 & 0.77 $\pm$ 0.14\\ 
  &&$Y$&--& -- & 0.94 & -0.47 $\pm$ 0.30\\ \cline{3-7}
  &&Combined$^{c}$ & 0.78 $\pm$ 0.04&1.28$^{+0.07}_{-0.06}$ \rule[-1mm]{0mm}{5mm}\\ \hline
\multicolumn {7} {l} {$^a$ The rms deviations of the post-fit residuals.}\\
\multicolumn {7} {l} {$^b$ 1 mas yr$^{-1}$ = 6.1 km s$^{-1}$ at the distance.}\\
\multicolumn {7} {l} {$^c$ The weighted means of those for two features.} \\
     \end{tabular}
\end{table*}

\begin{table*}
 \centering
  \caption{Gaussian Parameters of $\pi$ distributions from Bootstrap Analysis}\label{tb:bootstrap}
   \begin{tabular}{cccccc}\hline
  Feature&Median (mas)&Standard Deviation &D (kpc)&3-$\sigma$$^{a}$ &\\ \hline \hline
  O1&0.80 $\pm$ 0.005& 0.06 $\pm$ 0.005&1.25$^{+0.11}_{-0.09}$&$^{+0.38}_{-0.24}$\rule[-0.5mm]{0mm}{4mm}\\
  C6&0.75 $\pm$ 0.008& 0.10 $\pm$ 0.008&1.33$^{+0.20}_{-0.15}$&$^{+0.87}_{-0.38}$\rule[0mm]{0mm}{4mm}\\ \hline
  Combined$^{b}$ & 0.78 & 0.05&1.28$^{+0.09}_{-0.08}$&$^{+0.33}_{-0.22}$ \rule[-1mm]{0mm}{5mm}\\ \hline
\multicolumn {6} {l} {$^a$ 3-$\sigma$ ($>$ 99 $\%$ confidence) error ranges in kpc.} \\
\multicolumn {6} {l} {$^b$ The weighted means of those for two features.} \\
     \end{tabular}
\end{table*}

  \subsection{Astrometric Error Sources}
 The post-fit residuals seen in our parallax and proper motion measurement 
 are much larger than the formal errors of the elliptical Gaussian fittings. 
 This can be seen especially in $Y$ fitting, where residual variance is about 850 $\mu$as. 
 We again note that the parallax in $Y$ degenerated into negligible order because of the source location. 
 This fact suggests that $Y$ fluctuations from a simple linear motion directly 
 reflect a degree of actual error in each position measurement. 
 One of the main source of this error is thought to be atmospheric zenith delay residuals 
 as discussed in several studies with VERA (e.g., Sato et al. 2008 and references therein). 
 We first check this in our case using the several expressions in \citet{Nakagawa2008}. 
 
 They assumed the error in station positions and image quality of phase-referenced map as additional error sources. 
 The former is only about 7 $\mu$as with the separation angle of 1.92$^{\circ}$ and baseline error of 3 mm which is the typical value for VERA. 
 This is, of course, negligible in our case. 
 The latter can be expressed by a relevant beam size in $X$ and $Y$ direction divided by a dynamic range of image. 
 The beam position angle are almost parallel to the $Y$ direction in our observation, and then, 
 expected error is 80 $\mu$as in $X$ and 200 $\mu$as in $Y$. 
 On the other hand, an error caused by atmospheric zenith delay residual is proportional 
 to a difference of signal path length between 2 beams divided by a relevant baseline length. 
 This can be calculated from an atmospheric zenith delay residual, 
 typical zenith angle and separation angle of the sources (see equation (1) in Nakagawa et al. 2008). 
 If we adopt a residual of 3 cm as a upper limit in our case and average zenith angle of 65$^{\circ}$, 
 atmospheric error is about 330 $\mu$as in $X$ and 820 $\mu$as in $Y$. 
 This is almost consistent with the result of Monte Calro Simulations in \citet{Honma2008b}. 
 
 Because J1755-2232 looks an ideal point source, the internal structure of J1755-2232 does not cause significant error. 
 If we also assume that the effect of internal structres of maser features is also negligible for simplicity, 
 we finally gets the total error of 340 and 850 $\mu$as in each direction from the root sum square of these 3 values.
 As one can see, this is consistent with the standard deviation of the post-fit residuals (see table \ref{tb:parallax}). 
 If differences of a separation angle, pair position angle and zenith delay residuals are taken into account, 
 this can be roughly comparable to the value in \citet{Choi2008}, where the source declination is almost same in our case. 
 However, the expected error of the parallax is estimated to be $\sim$ 0.09 mas in a statistical way, here, 
 we divide 340 $\mu$as by $\sqrt{2\times(8-1)}$ considering 8 observing epochs and 2 distinct maser features. 
 This is clearly larger than the actual fitting errors of least square analysis (0.04 mas). 
 
 This seems to indicate that the effect of the variability of maser 
 feature structures, which we have ignored above, cannot be negligible. 
 This type of error is another important error source for VERA observations (e.g., \cite{Hirota2007}).  
 Moreover, it can be actually dominant error source in some cases (e.g., \cite{Sato2008}). 
 If we attribute dominant part of the post-fit variance to this effect, instead of the atmospheric error, 
 expected error of the parallax can be reduced by the additional factor of $1/\sqrt{N}$, here $N$ is the number of maser spots in single feature. 
 Since both of the feature O1 and C6 are consist of 7 maser spots in our case, 
 the conclusive error of parallax is to be $0.33/\sqrt{7\times 2\times (8-1)}$ $\sim$ 0.03 mas, 
 where the numerator of 0.33 mas is averaged value of the post-fit residual in $X$. 
 This is just comparable to the actual fitting error. 
 
 Therefore, we conclude that the dominant error source is the variability of maser feature structures in our case, 
 although the contribution from atmospheric zenith delay residuals is also nonnegligible. 
 The proportion of their contributions is, more quantitatively, to be $\sim$ 2:1, 
 if we use averaged atmospheric delay residuals of $\sim$ 1.5 cm which has been evaluated from the image optimizing method. 
 
 \subsection{Kinematics and Spatial Distribution of Masers}
 Absolute proper motions of feature O1 and C6 shown in table \ref{tb:parallax} are motions respect to the Sun, 
 and hence, include the contribution of solar motion and the Galactic rotation. 
 If we assume the solar motion relative to the LSR based on the Hipparcos data \citep{Dehnen1998}, 
 the contribution of solar motion is calculated to be 0.69 mas yr$^{-1}$ and -1.19 mas yr$^{-1}$ for $X$ and $Y$, respectively. 
 The contribution of the Galactic rotation is estimated to 
 be -0.04 mas yr$^{-1}$ and -0.07 mas yr$^{-1}$ based on a $R_{0}$ of 8.4 kpc and $\Theta_{0}$ of 254 km s$^{-1}$ \citep{Reid2009}. 
 This value is well negligible and do not change in the common range of $R_{0}$ and $\Theta_{0}$ (220 -- 254 km s$^{-1}$, 8.0 -- 8.5 kpc;\cite{Hou2009}). 
 Here, we assumed flat rotation and adopted our newly determined distance of 1.28 kpc. 
 The intrinsic proper motions of the maser features, where these two contributions are subtracted, is represented in table \ref{tb:prop}. 
 We note that these motions are still include the peculiar motion of their natal cloud and the internal motions of each maser feature.
 Since only two features are detected in 2--beam analysis, we cannot divide these two components at this section, 
 but brief presumption will be provided in section 5.2.2 based on the model fitting of the maser kinematics. 
 
\begin{figure}
  \begin{center}
    \FigureFile(80mm,150mm){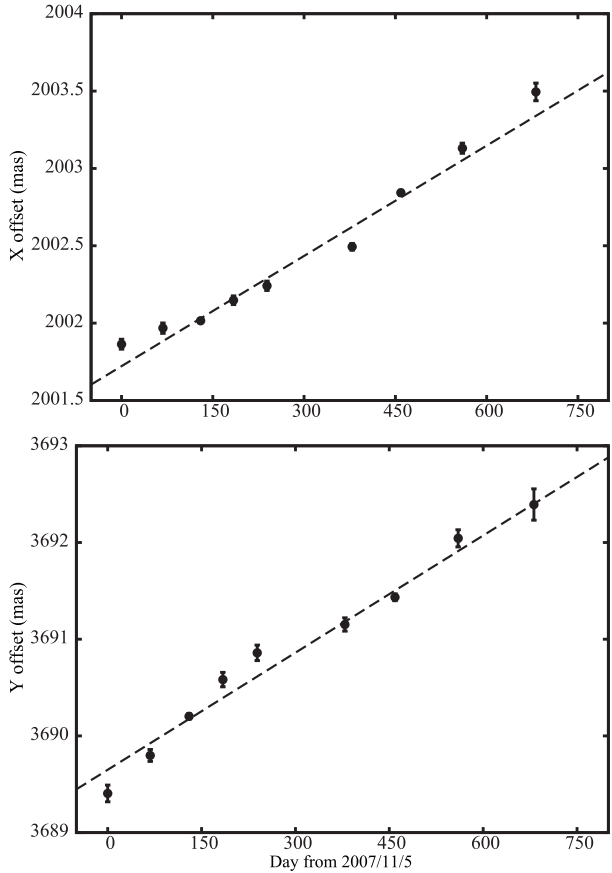}
  \end{center}
  \caption{The internal proper motion of feature C6. 
  $Y$ axes in each panel show the positional offset relative to the phase-referenced feature O1. 
  Both of Dashed lines indicate the best-fit linear proper motions. }\label{fig:inmotion}
\end{figure}

\begin{figure*}
  \begin{center}
    \FigureFile(130mm,50mm){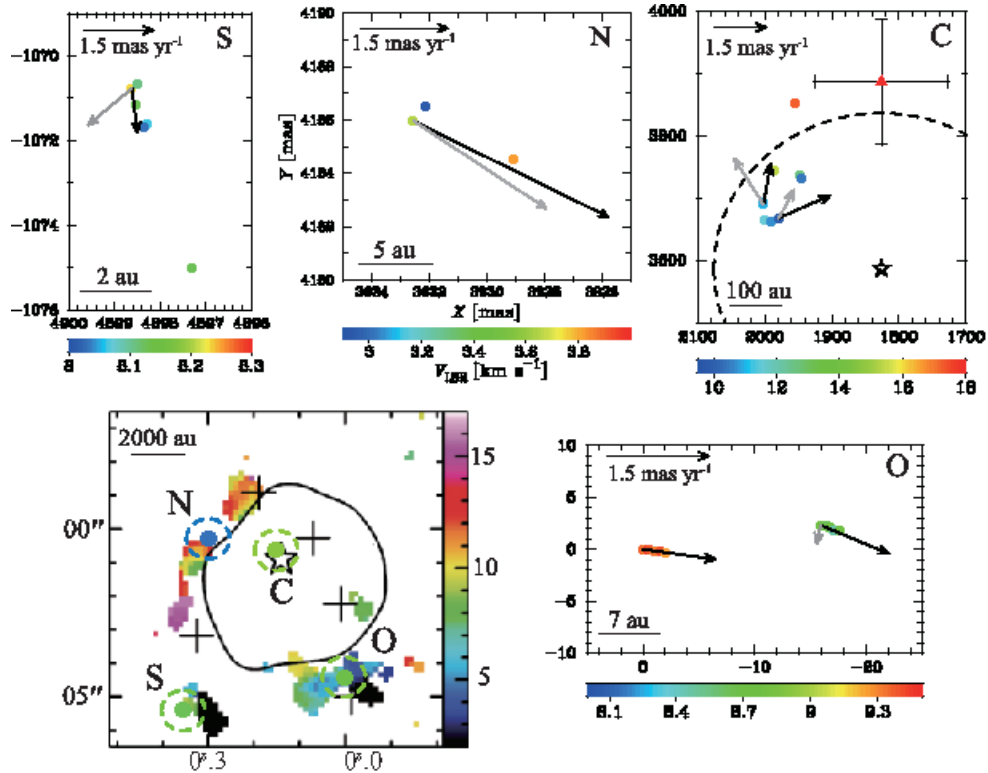}
  \end{center}
  \caption{Overall distribution of maser features. 
  Lower left panel: The locations of 4 maser site (O,C,N,S) are presented by color points with dashed circle. 
  The coordinate origin is $\alpha_{2000}$=\timeform{18h00m30.3s}, $\delta_{2000}$=\timeform{-24D04'0.0"}. 
  Here, background color scale, black crosses and contour shows the first moment map of SiO $J$ = 8--7 emission, 
  875 $\mu$m dust cores and 3 mm continuum image (0.1 Jy beam$^{-1}$) from \citet{Hunter2008}, respectively. 
  The star marks the positions of Feldt's star $\alpha_{2000}$=\timeform{18h00m30.44s}, $\delta_{2000}$=\timeform{-24D04'0.9"}. 
  Other 4 panels: Detailed distributions of maser features are shown. 
  Here, each point indicates detected maser feature. 
  Each axis shows the relative coordinate from the phase-referenced maser feature O1 in the units of mas. 
  Absolute and internal proper motions are written in black and grey arrows, respectively. 
  Red triangle with error bar in site C (upper right panel) is the maser detected by \citet{Hofner1996}. 
  Here, the star marks the position of Feldt's star again and its error is presented by dashed circle of 250 mas diameter 
  (200 from positional error and 50 from PSF, see \cite{Feldt2003}). 
  The color represents $V_{\mathrm{LSR}}$ (km s$^{-1}$) for all of 5 panels in the scale of associated color bars. }\label{fig:Feature}
\end{figure*}

 Additional 12 maser features were detected in the single beam analysis. 
 All parameters of detected features are listed in table \ref{tb:Feature}. 
 Total 14 maser features can be divided into 4 maser sites. 
 We name these sites as origin (O), center (C), north (N) and south (S) based on their positions. 
 These four maser sites are widely spread around the UCH$\emissiontype{II}$ region. 
 We estimated internal proper motions relative to the phase-referenced feature O1 
 for the features which were detected in at least three observing epochs. 
 Figure \ref{fig:inmotion} shows the example of feature C6. 
 Their internal motion was well fitted by linear motion. 
 Estimated motions are also summarized in table \ref{tb:prop}. 
 Overall distributions and proper motions of maser features are represented in figure \ref{fig:Feature}, 
 where black and grey arrows show relative and converted absolute proper motions, respectively. 
 Relative positions and internal proper motions of all maser features are 
 converted to absolute values using the position and motion of the feature O1. 
 The converted motion of C6 is actually coincide with the directly estimated one within the error. 
 
 Each of the sites O, N and S, which is located on outside of the ionized shell, contains only one or two maser features. 
 This limited number of maser features can be related to a life time of H$_{2}$O maser activity. 
 They are generally believed to disappear along with a evolution of UCH$\emissiontype{II}$ region (e.g., \cite{Beuther2002b}, \cite{Breen2010}), 
 in attributing to a dispersion of dense gas which required to excite masers. 
 All of these sites are associated with SiO $J$ = 8 -- 7 emission 
 detected by Sub-Millimeter Array (SMA) \citep{Hunter2008}. 
 SiO $J$ = 8 -- 7 emission is a moderate shock tracer and frequently observed in 
 post-shock gas which is associated with a protostellar outflow (e.g., \cite{Takami2006}). 
 Such a spatial relationship may indicate that H$_{2}$O maser and SiO emission trace same shock fronts. 
 This is not so surprising case because H$_{2}$O masers are also associated with strong outflow shock frequently (e.g., \cite{Motogi2008}; \cite{Torrelles2010}). 
 
 On the other hands, line of sight velocities of maser features in the site O and N are not consistent with that of SiO emission. 
 The velocity offsets are larger than 5 km s$^{-1}$. 
 These offsets may reflect the difference of precise locations where each emission comes from. 
 This can be simply attributed to the different excitation conditions of SiO and H$_{2}$O. 
 The critical density of SiO $J$ = 8 -- 7 emission are $n_{\mathrm{H_{2}}}$ $\cong$ 10$^{7}$ cm$^{-3}$ based on the database in \citet{Schoier2005}. 
 It is actually 2 orders of magnitude smaller than that of H$_{2}$O maser and comparable with pre-shock density of the maser \citep{Elitzur1992}. 
 H$_{2}$O maser is, hence, probably excited in the most strongly compressed part such as a head part of a bow shock. 
 Emission from that region cannot dominate the integrated SiO emission, 
 since such a region should have quite limited volume compared to total post-shock gas. 
 
 The site C, by contrast to other site, is not associated with SiO emission. 
 There are 9 maser features and almost all of them (7 of 9) are highly variable and only detected in a single epoch. 
 This maser site had been reported in the past VLA observation \citep{Hofner1996}, 
 but observed position and line of sight velocity was slightly different from our detection. 
 Whole maser features, including the one detected by VLA, closely located at the position of Feldt's star ($<$ 200 mas). 
 Their line of sight velocities are little red-shifted from systemic velocity of 9 km s$^{-1}$ (e.g., \cite{Hunter2008}) 
 and clearly different from that of OH masers seen in the same position \citep{Stark2007}. 
 
 Highly blue shifted ($\sim$ -35 km s$^{-1}$) OH masers are thought to be excited in expanding neutral shell 
 or strong outflow (the former in \cite{Stark2007}, the latter in \cite{Zijlstra1990}). 
 They seem to be located on the foreground of the ionized shell in either cases. 
 The kinematic difference strongly invokes the distinct origins of OH and H$_{2}$O masers. 
 Such a nearly located, but strictly distinct displacement of these two maser was quite natural 
 if we considered different excitation conditions again, and it had been actually observed for several regions \citep{Forster1989}. 
 Taking into account these situation, we propose that this H$_{2}$O maser site is just located inside a ionized shell and 
 excited in a remnant of dense circumstellar structure such as an accretion disk. 
 Extremely young dynamical age of UCH$\emissiontype{II}$ region ($\sim$ 600 yr) well support the presence of such a remnant and 
 their crescent-like distribution and velocity field can be actually explained by a simple ring model (see section 5.2.2). 
 
 \begin{table*}
 \centering
  \caption{Intrinsic Proper Motions}\label{tb:prop}
   \begin{tabular}{ccccc}\hline
   Feature & \multicolumn{2}{c}{Absolute Proper Motion$^{a}$} &\multicolumn{2}{c}{Relative Proper Motion$^{a}$}\rule[0mm]{0mm}{2mm}\\
   & $\mu_{X}$  &$\mu_{Y}$ & $\mu_{X}$ & $\mu_{Y}$ \rule[-2mm]{0mm}{0mm}\\ \hline
O1$^{b}$ & -1.08 $\pm$ 0.24 & -0.16 $\pm$ 0.40& -- & -- \\
C6 & 0.12 $\pm$ 0.29 & 0.79 $\pm$ 0.47 & -- & -- \\ \hline
O2$^{c}$ & -1.02 $\pm$ 0.33 & -0.46 $\pm$ 0.43 & 0.06 $\pm$ 0.09 & -0.29 $\pm$ 0.03 \\
C6 & -0.21 $\pm$ 0.29 & 1.31 $\pm$ 0.48 & 0.87 $\pm$ 0.05 & 1.47 $\pm$ 0.08 \\
C9 & -1.51 $\pm$ 0.27 & 0.72 $\pm$ 0.53 & -0.43 $\pm$ 0.03 & 0.89 $\pm$ 0.13 \\
N & -3.38  $\pm$ 0.43 & -1.79 $\pm$ 1.56 & -2.3 $\pm$ 0.19 & -1.63 $\pm$ 1.16 \\
S1 & -0.12 $\pm$ 0.40 & -1.03 $\pm$ 0.93 & 0.96 $\pm$ 0.16 & -0.86 $\pm$ 0.53 \\ \hline
\multicolumn {5} {l} {$^a$ All values are in the units of mas yr$^{-1}$.}\\
\multicolumn {5} {l} {$^b$ The upper 2 rows are from the parallax fitting.}\\
\multicolumn {5} {l} {$^c$ The lower 5 rows are from the internal proper motions relative to O1.}\\
     \end{tabular}
\end{table*}
 
 \begin{table*}
 \centering
  \caption{Detected Maser Features}\label{tb:Feature}
   \begin{tabular}{ccccccc}\hline
   Feature & Epoch & $V_{\mathrm{LSR}}$& $\Delta$$V^{a}$ & \multicolumn{2}{c}{Offset$^{b}$ (mas)} & Peak Intensity \rule[0mm]{0mm}{2mm}\\
   & & \multicolumn {2}{c}{(km s$^{-1}$)}& $X$ (err) & $Y$ (err) & (Jy beam$^{-1}$)\rule[-2mm]{0mm}{0mm}\\ \hline
O1 & 1 & 9.36  & 1.26  & 0.00  (0.03)  & 0.00  (0.08)  & 74.0  \\
& 2 & 9.34  & 1.69  & -0.20  (0.01)  & -0.03  (0.02)  & 75.6  \\
& 3 & 9.26  & 1.90  & -0.38  (0.01)  & -0.06  (0.01)  & 80.3  \\
& 4 & 9.50  & 2.11  & -0.54  (0.01)  & -0.08  (0.02)  & 72.1  \\
& 5 & 9.37  & 1.47  & -0.71  (0.02)  & -0.11  (0.04)  & 63.6  \\
& 6 & 9.33  & 1.26  & -1.12  (0.02)  & -0.17  (0.05)  & 32.0  \\
& 7 & 9.30  & 1.26  & -1.36  (0.01)  & -0.21  (0.02)  & 19.6  \\
& 8 & 9.39  & 1.05  & -1.65  (0.04)  & -0.25  (0.11)  & 21.7  \\
& 9 & 9.26  & 1.26  & -2.01  (0.04)  & -0.31  (0.11)  & 14.2  \\
O2 & 1 & 8.72  & 0.63  & -15.90  (0.06)  & 2.33  (0.17)  & 10.1  \\
& 2 & 8.70  & 0.84  & -16.24  (0.06)  & 2.30  (0.10)  & 5.5  \\
& 3 & 8.63  & 1.05  & -16.51  (0.04)  & 2.25  (0.10)  & 7.0  \\
& 4 & 8.65  & 0.84  & -16.63  (0.06)  & 2.12  (0.15)  & 5.9  \\
& 5 & 8.74  & 0.63  & -16.85  (0.07)  & 2.08  (0.17)  & 5.1  \\
& 6 & 8.48  & 0.63  & -17.16  (0.08)  & 1.87  (0.25)  & 3.4  \\
& 7 & 8.46  & 0.84  & -17.20  (0.06)  & 1.76  (0.14)  & 3.2  \\
& 8 & 8.76  & 0.63  & -17.69  (0.09)  & 1.78  (0.23)  & 2.7  \\ \hline
C1 & 8 & 17.19  & 0.63  & 1954.03  (0.05)  & 3852.79  (0.14)  & 5.1  \\
C2 & 2 & 15.45  & 0.63  & 1984.14  (0.07)  & 3745.18  (0.12)  & 4.2  \\
C3 & 7 & 12.67  & 1.69  & 1944.75  (0.04)  & 3732.56  (0.08)  & 3.3  \\
C4 & 7 & 12.67  & 0.84  & 1946.13  (0.09)  & 3736.36  (0.21)  & 1.3  \\
C5 & 3 & 11.58  & 0.63  & 1998.60  (0.07)  & 3665.11  (0.17)  & 2.8  \\
C6 & 1 & 11.04  & 1.05  & 2001.86  (0.03)  & 3689.41  (0.09)  & 27.9 \\
& 2 & 11.02  & 1.26  & 2001.77  (0.03)  & 3689.77  (0.06)  & 19.8  \\
& 3 & 11.16  & 1.47  & 2001.63  (0.01)  & 3690.14  (0.03)  & 43.0  \\
& 4 & 11.18  & 1.68  & 2001.60  (0.03)  & 3690.50  (0.07)  & 56.0  \\
& 5 & 11.27  & 1.26  & 2001.54  (0.03)  & 3690.75  (0.08)  & 46.9  \\
& 6 & 11.22  & 1.26  & 2001.37  (0.02)  & 3690.98  (0.07)  & 39.8  \\
& 7 & 11.19  & 1.47  & 2001.49  (0.02)  & 3691.23  (0.04)  & 28.3  \\
& 8 & 11.08  & 1.47  & 2001.48  (0.03)  & 3691.79  (0.09)  & 13.7  \\
& 9 & 10.94  & 1.05  & 2001.48  (0.06)  & 3692.09  (0.16)  & 4.7  \\
C7 & 7 & 10.56  & 0.63  & 1944.66  (0.09)  & 3732.89  (0.21)  & 1.9  \\
C8 & 3 & 10.74  & 0.63  & 1990.50  (0.05)  & 3662.93  (0.11)  & 5.1  \\
& 1 & 9.78  & 0.63  & 1978.74  (0.03)  & 3668.10  (0.08)  & 22.1  \\
& 2 & 9.76  & 0.84  & 1978.48  (0.06)  & 3668.15  (0.10)  & 9.9  \\
& 3 & 9.90  & 0.84  & 1978.20  (0.03)  & 3668.34  (0.08)  & 8.2  \\
& 5 & 10.00  & 0.63  & 1977.75  (0.08)  & 3668.60  (0.20)  & 3.9  \\ \hline
N & 2 & 3.65  & 1.05  & 3932.54  (0.05)  & 4185.95  (0.09)  & 6.7  \\
& 3 & 2.95  & 2.95  & 3932.14  (0.02)  & 4186.54  (0.04)  & 18.0 \\
& 7 & 3.82  & 1.26  & 3929.04  (0.05)  & 4184.55  (0.11)  & 2.9  \\ \hline
S1 & 4 & 8.23  & 1.26  & 4898.62  (0.07)  & -1070.77  (0.19)  & 8.0  \\
& 5 & 8.11  & 1.05  & 4898.48  (0.04)  & -1070.66  (0.10)  & 13.7  \\
& 6 & 8.06  & 1.47  & 4898.27  (0.05)  & -1071.59  (0.15)  & 12.6  \\
& 7 & 8.03  & 1.47  & 4898.35  (0.03)  & -1071.68  (0.07)  & 11.3  \\
& 8 & 8.13  & 1.05  & 4898.50  (0.05)  & -1071.14  (0.14)  & 4.7   \\
S2 & 8 & 8.13  & 0.63  & 4897.28  (0.09)  & -1075.02  (0.24)  & 2.4 \\ \hline
\multicolumn {5} {l} {$^a$ Full width at zero intensity (FWZI) for each maser feature.}\\
\multicolumn {5} {l} {$^b$ The positions relative to (\timeform{18h00m30.3066s}, \timeform{-24D04'4.48649"}) (J2000.0).}\\
     \end{tabular}
\end{table*}

 \section{Discussion}
 \subsection{Validity of New Distance}
 The newly estimated distance of 1.28 kpc is well smaller than the previously reported distances ($>$ 1.9 kpc). 
 This is mainly because almost all of past measurements have been done kinematically. 
 Kinematic distance essentially depends on the Galactic rotation model and accuracy of determining rotation velocity of the source. 
 If it is the case that target source has unknown, non-negligible velocity component aside from the Galactic rotation, 
 kinematic distance is no longer reliable in principle. 
 This type of systematic error is more and more significant especially for the source near the Galactic center direction. 
 In the case of our source ($l$ = 5$^{\circ}$.89), only a few km s$^{-1}$ offset along the line of sight can cause an error of $\sim$ 1 kpc. 
 
 \citet{Reid2009} had first analyzed the Galactic rotation based on 
 the distances of 18 star-forming regions which are estimated only from trigonometric parallaxes of masers derived from VLBA and VERA observations. 
 Their new analysis indicates that commonly used kinematic distances are generally overestimated, sometimes by factors greater than 2, which is just in our case. 
 Similar result has already been reported in other VERA observation (G14.33-0.64, the measured distance $\sim$ 1.1 kpc, $l$ = 14$^{\circ}$.33 \citep{Sato2010a}). 
 Our result seems to be well consistent with the hypothesis in \citet{Sato2010a} that the Sagittarius spiral arm, 
 in which G5.89-0.39 and G14.33-0.64 are thought to be located, lies at the closer distance ($\sim$ 1 kpc) from the sun compared to the previous value (2 -- 3 kpc). 
 Of course, more large samples must be required to confirm it. 
 
 Progressive Galactic simulation in \citet{Baba2009} also suggests that star-forming regions and young stars in a spiral arm 
 have significant non-circular motions up to 30 km s$^{-1}$. 
 If their calculation is correct, a kinematic distance intrinsically has large systematic error of $\sim$ 2--3 kpc. 
 With these contexts, the significant difference between the kinematic distance and our new distance can be 
 attributed to the systematic error in the kinematic distance. 
 
 ACW98 have been derived the source distance of 2.0 kpc from the expansion of the ionized shell, 
 but it is still highly model dependent (see section 5.2.3). 
 Consequently, our new distance which has no model dependency is thought to be the most reliable distance of G5.89-0.39. 
 
 \subsection{Physical Parameters and High Mass Star Formation in G5.89-0.39}
 Because of significant modification of the distance to G5.89-0.39, 
 we first recalculated physical parameters which had been previously reported based on the new distance. 
 Table \ref{tb:Ostar} to \ref{tb:Ioutflow} summarize recalculated parameters. 
 Each table contains both of original and modified values with the reference papers. 
 Detailed explanations for listed parameters are given in related subsections and captions. 
 
\begin{table*}
 \centering
  \caption{Recalculated Stellar Parameters}\label{tb:Ostar}
   \begin{tabular}{l|c|c|c}\hline
   Parameter&Original value&Modified value&Reference\\ \hline
   $N_{\mathrm{L}}$ (s$^{-1}$)& 4.5$\times$10$^{48}$ &1.6$\times$10$^{48}$ &(1)\\
   $L_{\mathrm{FIR}}$ (\LO)& 3.0$\times$10$^{5}$ &7.3$\times$10$^{4}$ &(1)\\
   Sp Type (radio) & O7 &O8.5&(1), (2)\\
   Sp Type (FIR) & O6 &O8&(1), (2)\\
   $M_{\mathrm{*}}$ (\MO)& $\sim$ 40 &$\sim$25 &(3)\\
   $\dot{P}_{\mathrm{wind}}$ (\MO km s$^{-1}$ yr$^{-1}$ )& -- &2.5 $\times$ 10$^{-3}$ &(4)\\ \hline
\multicolumn {4} {l} {References.-- (1):\cite{Wood1989}, (2):\cite{Panagia1973},}\\
\multicolumn {4} {l} {(3):\cite{Vacca1996}, (4):this work.}\\
     \end{tabular}
\end{table*}

 \begin{table*}
 \centering
  \caption{Recalculated Parameters of UCH$\emissiontype{II}$ region}\label{tb:UCHIIregion}
   \begin{tabular}{l|c|c|c}\hline
   Parameter&Original value&Modified value&Reference$^{a}$\\ \hline
   $n_{\mathrm{e}}$ (cm$^{-3}$)& 2.4$\times$10$^{5}$ &3.2$\times$10$^{5}$ &(1)\\
   $R_{\mathrm{i}} / R_{\mathrm{e}}$ (au)& 1900 / 5200&1200 / 3300 &(2)\\
   $\upsilon_{\mathrm{shell}}$$^{b}$ (km s$^{-1}$)& 39 &25  &(2)\\
   $t_{\mathrm{dyn}}$ (yr)& 600 & 600 &(2)\\ 
   $E_{\mathrm{Therm}}$ (10$^{45}$ erg)& -- &0.7 &(3)\\
   $E_{\mathrm{Kin}}$ (10$^{45}$ erg)& -- &1.2 &(3) \\
   $P_{\mathrm{shell}}$ (\MO km s$^{-1}$ )& -- &5.6 &(3)\\ \hline
   $M_{\mathrm{env}}$ (\MO)& 300 &123 &(4)\\
   $n_{\mathrm{env}}$ (cm$^{-3}$)& 5.3 $\times 10^{6}$ &8.0 $\times 10^{6}$ &(4)\\
   $T_{\mathrm{env}}$ (K)& 40 -- 140 & 40 -- 140 &(5)\\ \hline
\multicolumn {4} {l} {$^a$References.-- (1):\cite{Wood1989}, (2):ACW98, (3):this work, }\\
\multicolumn {4} {l} {(4):\cite{Tang2009}, (5):\cite{Su2009}.}\\
\multicolumn {4} {l} {$^b$ Tangential expansion derived from the angular expansion rate.}\\
     \end{tabular}
\end{table*}

\begin{table*}
 \centering
  \caption{Parameters of the E-W Outflow (CO $J$ =1 -- 0)) from \citet{Watson2007}}\label{tb:Ooutflow}
   \begin{tabular}{l|cc|cc}\hline
   Parameter& \multicolumn {2} {c|} {Original value}& \multicolumn {2} {c} {Modified value} \\ \hline
   $t_{\mathrm{dyn}}$ (yr) & \multicolumn {2}{c|}{7700}&\multicolumn {2}{c}{5000} \\ \hline
   &Blue lobe&Red lobe&Blue lobe&Red lobe\\ \hline
   $\upsilon_{\mathrm{flow}}^{a}$ (km s$^{-1}$) & --15.0 -- 3.2 &13.2 -- 25.0 & --15.0 -- 3.2 &13.2 -- 25.0\\ 
   $M$ (\MO)& 123 & 116 &50.4& 47.5 \\
   $E_{\mathrm{kin}}$ ($10^{46}$ erg)& 6.4 &5.3 &2.6 & 2.2 \\
   $P$ (10$^{2}$ \MO km s$^{-1}$)& 9.0& 8.0&3.7 & 3.3 \\ 
   $\dot{M}$ (10$^{-3}$ \MO yr$^{-1}$)& 16.0& 15.0&10.0 &9.6 \\
   $\dot{P}$ (10$^{-2}$ \MO km s$^{-1}$ yr$^{-1}$ )& 12.0 & 10.0 &7.7 &6.4 \\
   $L_{\mathrm{mech}}$ (\LO)& 68.0& 56.0&44.0&36.0\\ \hline
\multicolumn {5} {l} {$^a$ Integrated velocity range.}\\
     \end{tabular}
\end{table*}

\begin{table*}
 \centering
  \caption{Parameters of the inner high velocity Outflow (CO $J$ =3 -- 2)) from \citet{Klaassen2006}}\label{tb:Ioutflow}
   \begin{tabular}{l|cc|cc}\hline
   Parameter& \multicolumn {2}{c|}{Original value$^{a}$}& \multicolumn {2}{c}{Modified value} \\ \hline
   $t_{\mathrm{dyn}}$ (yr) & \multicolumn {2}{c|}{2000}&\multicolumn {2}{c}{1300} \\ \hline
   &Blue lobe&Red lobe$^{c}$&Blue lobe&Red lobe$^{c}$\\ \hline
   $\upsilon_{\mathrm{flow}}^{b}$ (km s$^{-1}$) & -66.0 &78.0 &-66.0 & 78.0\\ 
   $M$ (\MO)& 2.7& 0.6&1.1& 0.25 \\
   $E_{\mathrm{kin}}$ ($10^{46}$ erg)& 1.5& 0.3 &0.6 & 0.1 \\
   $P$ (\MO km s$^{-1}$)& 55.9& 12.0&22.0 & 4.9 \\ 
   $\dot{M}$ ($10^{-4}$ \MO yr$^{-1}$)& 13.5& 3.0 &8.5 &1.9 \\
   $\dot{P}$ ($10^{-2}$ \MO km s$^{-1}$ yr$^{-1}$ )& 2.8 & 0.6&2.7 &0.4 \\
   $L_{\mathrm{mech}}$ (\LO)& 59.0& 11.5 &38.0 &7.4\\ \hline
\multicolumn {5} {l} {$^a$ Inclination of outflow axis which has been originally assumed as 45$^{\circ}$ is ignored}\\
\multicolumn {5} {l} {$^b$ The maximum velocities.}\\
\multicolumn {5} {l} {$^c$ Absorption features seen in the red lobe have not been corrected }\\
\multicolumn {5} {l} {in \citet{Klaassen2006}.}\\
     \end{tabular}
\end{table*}

 \subsubsection{Evolution of the O-type Protostar}
 Table \ref{tb:Ostar} show the properties of Feldt's star. 
 We derived stellar spectral type in the same manner in \citet{Wood1989}. 
 We first estimated excitation parameter of the source from radio continuum emission (see also \cite{Kurtz1994}). 
 This was simply converted to the flux of Lyman continuum photons ($N_{\mathrm{L}}$) and spectral type (Sp type)
 comparing the stellar models in \citet{Panagia1973}. 
 In the second, we used total far-infrared luminosity ($L_{\mathrm{FIR}}$).
 $L_{\mathrm{FIR}}$ can be a good measure of bolometric luminosity for deeply embedded sources. 
 Because of the low spatial resolution, $L_{\mathrm{FIR}}$ may include several contributions 
 from other objects around the UCH$\emissiontype{II}$ region, and hence, this gives upper limit of the source luminosity. 
 
 Derived spectral types are almost consistent and indicate late O-type star (O8 -- O8.5 ZAMS). 
 Expected stellar mass $M_{\mathrm{*}}$ is about $\sim$25 $M_{\odot}$ (e.g., \cite{Vacca1996}). 
 This is a roughly half value of previously reported O5 ZAMS (or O5V) star. 
 \citet{Feldt2003} tested such an early spectral type by a model fitting of mid-infrared color ($K_{s}$ and $L^{\prime }$ bands). 
 But their fitting is based on the distance of 1.9 kpc, and if we adopt new distance and appropriate extinction value, 
 all of O type stellar models which are later than O5 can move into the range of $L^{\prime }$ band excess reported in \citet{Feldt2003}. 
 
 The line ratio of He$\emissiontype{I}$ triplet at 2.11 $\mu$m to Br$\gamma$ also supports the existence of late O-type star. 
 \citet{Hanson2002} have observationally determined upper limit of this ratio as 0.02 for G5.89-0.39 
 and indicated that this value is clearly smaller than a value expected for a star earlier than O7 type. 
 \citet{Puga2006} actually detected 2.11 $\mu$m He$\emissiontype{I}$ emission with VLT. 
 Although they suggested rather early O star ($<$ O7), 
 the detection only at the location of Feldt's star also ruled out a star earlier than O7, 
 because such an early type star could fully ionize whole He atoms and provide constant line ratio throughout a UCH$\emissiontype{II}$ region (e.g., \cite{Osterbrock1989}). 
 This seems to be a reliable constraint, since the line ratio is fully independent of the source distance. 
 Consequently, new spectral type of O8 -- O8.5 seems to be reasonable. 
 
 As seen in table \ref{tb:Ostar}, we also estimated the momentum rate of stellar wind ($\dot{P}_{\mathrm{wind}}$). 
 It has been calculated from pressure balance at the inner surface of ionized shell. 
 We simply assumed that spherically-symmetric wind and then 
 thermal pressure of ionized gas was balanced with $\dot{P}_{\mathrm{wind}}$ per unit area. 
 Although detailed properties of mass loss activity at the ZAMS stage are still unknown, 
 estimated $\dot{P}_{\mathrm{wind}}$ is ten times larger than that of a dwarf star which has $\sim$25 $M_{\odot}$ (\cite{Sternberg2003} and reference therein).
 A wind velocity is generally determined by gravitational force at the launching point of a wind, 
 and it is thought to be same order between ZAMS and dwarf stage. 
 This is about 2.0 $\times 10^{3}$ km s$^{-1}$ for 25 $M_{\odot}$ dwarf (e.g., \cite{Smith2002}). 
 In this case, mass loss rate is order of 10$^{-6}$ \MO$\:$ yr$^{-1}$, and of cource, ten times larger than the value of dwarfs . 
 One possible explanation for this excess may be larger stellar 
 radius of very young ZAMS star which is still under contraction.
 This could cause relatively weak surface gravity and might help mass loss activity.
 
 Recent theoretical works in \citet{Hosokawa2009} calculated detailed stellar evolution via spherically-symmetric mass accretion. 
 They have shown that large entropy supply under high accretion rates ($>$ 10$^{-4}$ $M_{\odot}$ yr$^{-1}$) onto a proto-stellar surface
 prevent contraction of protostar and, at some stage, cause far large stellar radius ($\sim$ 100 $R_{\odot}$).
 \citet{Hosokawa2010} has been treated the case of cold-disk accretion, and they have suggested that expected evolution is 
 qualitatively same as spherically-symmetric case after 10 $M_{\odot}$. 
 Realistic mass accretion should correspond to some intermediate state between these two extreme case as they pointed out. 
 
 The most important product of their evolutional track is delayed hydrogen burning. 
 Resultant young high mass (proto-)star with large radius and limited UV photons can explain highly luminous but low-effective temperature 
 source which has no detectable H$\emissiontype{II}$ region (e.g., \cite{Beuther2007}, \cite{Furuya2009}). 
 If Feldt's star has been just after the hydrogen ignition, 
 the maximum accretion rate, with which the star has been evolved, 
 is estimated to be $\sim$ 10$^{-3}$ $M_{\odot}$ yr$^{-1}$ from $\sim 25 M_{\odot}$ based on their calculations. 
 This assumption may mean that Feldt's star is quite rare sample, 
 but it seems to be well consistent with extraordinarily young dynamical age of UCH$\emissiontype{II}$ region. 
 If this is the case and we assume steady mass accretion, the time required to form Feldt's star is simply $\sim$ 2.5 $\times$ 10$^{4}$ yr. 
 
 \subsubsection{Maser Morphology Near the Feldt's Star: a Possible Partial Ring}
 Newly detected crescent-like structure of water maser may also suggest youth of the source. 
 As we mentioned above, the maser clumps in the site C can be associated with circumstellar remnant gas. 
 Their alignment and velocity field are actually fitted by the Keplerian rotating and expanding ring model in \citet{Uscanga2008}. 
 The fitting parameters are central position of a ring ($x_{0},y_{0}$), radius $R$, 
 inclination from the celestial plane $i$, position angle of apparent major axis measured from east toward north $\theta$, expansion (or infalling) velocity $\upsilon_{\mathrm{exp}}$ 
 and velocity of the source ($\upsilon_{x}$, $\upsilon_{y}$, $\upsilon_{z}$). 
 Here, $z$ means a line of sight direction. 
 We regarded Feldt's star as a central source of ring, in turn, its mass was fixed to be $25 M_{\odot}$. 
 Thus, Keplerian rotation velocity depended on $R$ only. 
 
\begin{figure*}
  \begin{center}
    \FigureFile(145mm,80mm){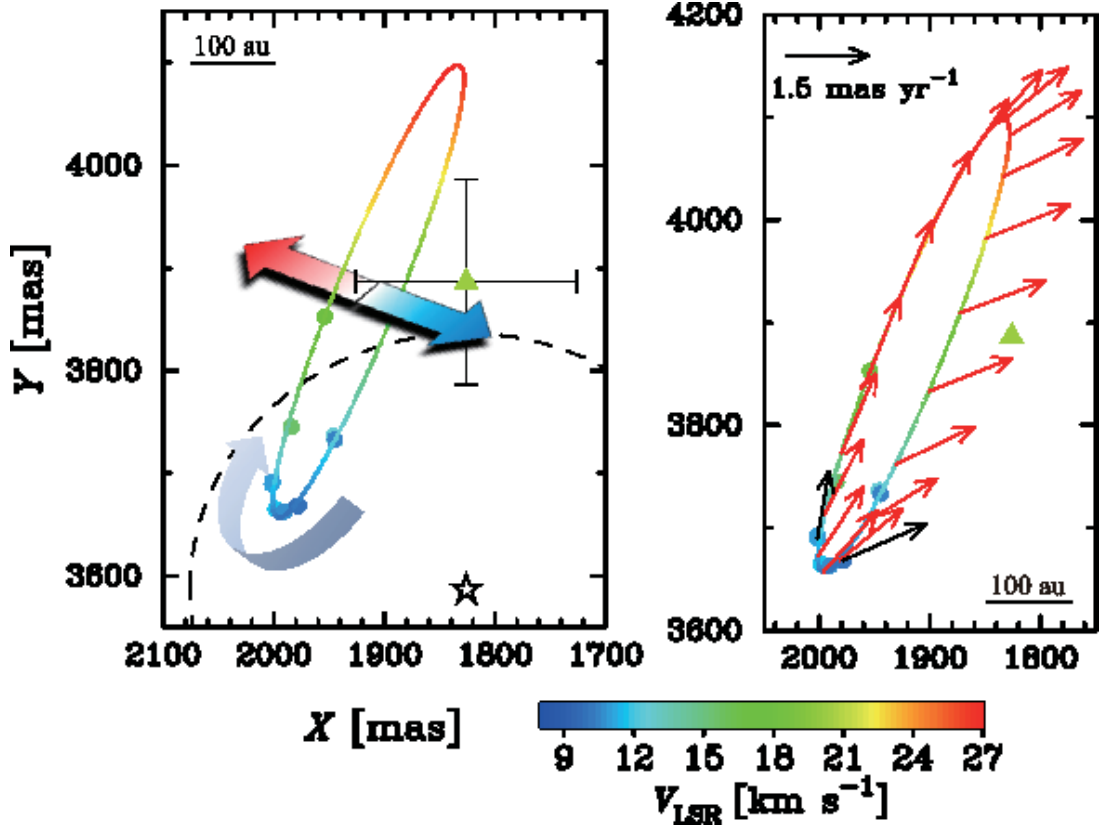}
  \end{center}
  \caption{Left panel shows the best-fit ring model superposed on the maser distribution in figure \ref{fig:Feature}. 
  Curved arrow represents rotational direction. 
  The red and blue arrows show a same geometry as red and blue lobe of E-W outflow. 
  The inclination of the ring is consistent with this outflow geometry. 
  Right panel shows velocity field of the ring. 
  Red and black arrows show expected proper motion vectors of the ring and observed vectors, respectively. 
  These motions contain both of the systemic velocity of G5.89-0.39 offset from the Galactic rotation 
  and migration velocity of the ring. }\label{fig:MaserRing}
\end{figure*}

 We performed the model fit with a standard $\chi^{2}$ fitting. 
 First, the spatial distribution of 9 (our detection) + 1 (VLA detection) maser features was fitted by $x_{0}$, $y_{0}$, $R$, $i$, $\theta$, and then, 
 $\upsilon_{x}$, $\upsilon_{y}$, $\upsilon_{z}$ were determined from the velocity field of masers. 
 It should be noted that $\upsilon_{x}$ and $\upsilon_{y}$ are more uncertain than $\upsilon_{z}$, 
 because there are only 2 available proper motion vectors. 
 Table \ref{tb:Ringfit} and figure \ref{fig:MaserRing} show the best fitted parameters and estimated ring superposed on the maser alignment, respectively. 

 The best fitted ring is almost edge-on and its position angle is roughly north -- sorth direction. 
 It is noteworthy that this geometry matches well the largest outflow which has a position angle of nearly linear alignment from east (red lobe) to west (blue lobe) (see below). 
 The center of the ring is slightly offset from Feldt's star (1826 mas, 3587 mas), but it can be within the ring if we take into account $\sim$ 200 mas error. 
 The small and negative expansion velocity means that the ring is slowly infalling. 
 
 Class $\emissiontype{II}$ CH$_{3}$OH masers which are believed to trace an accretion disk around HMPOs (e.g., \cite{Bartkiewicz2009} and reference therein),
 are not seen in G5.89-0.39 region. 
 It may be consistent with the fact that Class $\emissiontype{II}$ CH$_{3}$OH maser emission trace pre-UCH$\emissiontype{II}$ stage (e.g., \cite{Walsh1998}; \cite{Codella2000}). 
 On the other hand, a disk-like structure traced by 22 GHz H$_{2}$O masers has been actually reported in some sources. 
 \citet{Imai2006} has especially found infalling-rotating maser disk, which is similar case to ours, for early B-type HMPO. 
 \citet{Torrelles1996} have reported the H$_{2}$O maser disk candidate in Cepheus A HW2, which had almost same radius ($\sim$ 300 au) as our case. 
 If the crescent-like masers actually traces an accretion disk (or its remnant), it is clear indicative of a quite youthful O-type object 
 which is at least before complete disk evaporation. 
  
  \begin{table}
 \centering
  \caption{The best fit parameters of the ring model}\label{tb:Ringfit}
   \begin{tabular}{c|c}\hline
   Parameter& Best Fit (err$^{a}$)\\ \hline
   $x0$ (mas)& 1914.6 (0.5) \\
   $y0$ (mas)& 3877.2 (0.8) \\
   $R$ (mas) & 235.2 (1.2) \\
   $\theta$ ($^{\circ}$) & 110.5 (0.2) \\
   $i^{b}$ ($^{\circ}$) & +83.0 (0.2) \\
   $\upsilon_{\mathrm{exp}}$ (km s$^{-1}$)& -1.2 (0.8) \\
   $\upsilon_{\mathrm{x}}$ (km s$^{-1}$)& -6.5 (1.5) \\
   $\upsilon_{\mathrm{y}}$ (km s$^{-1}$)& 6.4 (1.1) \\
   $\upsilon_{\mathrm{z}}$ (km s$^{-1}$)& 18.6 (0.4) \\ \hline
\multicolumn {2} {l} {$^a$ Formal 4 $\sigma$ error of $\chi^{2}$ fittings.}\\
\multicolumn {2} {l} {$^b$ Western edge is on the far side.}\\
\end{tabular}
\end{table}

 We finally discuss about the velocity field. 
 The 3D motion of the ring ($\upsilon_{x}$, $\upsilon_{y}$, $\upsilon_{z}$) 
 still includes the non-circular (peculiar) motion of G5.89-0.39 system (recall section 4.3 and 5.1). 
 To divide these contributions, we now assume 2 conditions as follows, (1) Feldt's star has been initially located at the center of 
 the radio shell and migrated to the current position in the dynamical time of the shell, (2) the ring moved together with the star. 
 The possibility of stellar migration was already discussed in \citet{Feldt2003}. 
 This motion corresponds to be a proper motion of about 1.04 $\pm$ 0.30 and 0.98 $\pm$ 0.28 mas yr$^{-1}$ in $X$ and $Y$ direction, respectively. 
 By contrast, the migration velocity along a line of sight ($\sim$ 9.6 km s$^{-1}$) can be obtained from $\upsilon_{z}$ - $V_{\mathrm{LSR}}$. 
 The 3D motion of the partial ring can be eventually divided into the migration motion of Feldt's star and systemic peculiar motion of G5.89-0.39. 
 These divided values are presented in table \ref{tb:Vect}, where the line of sight component of the peculiar motion ($\sim$ 4.7 km s$^{-1}$) 
 was also extracted from $V_{\mathrm{LSR}}$ using the Galactic rotation model in \citet{Hou2009}. 
 
 \begin{table}
 \centering
  \caption{Divided velocity components}\label{tb:Vect}
   \begin{tabular}{c|ccc}\hline
   \multicolumn {4} {c} {Migration velocities of Feldt's star}\\ \hline
   ($\upsilon_{\mathrm{x}}$, $\upsilon_{\mathrm{y}}$, $\upsilon_{\mathrm{z}}$)&6.3$^{a}$ (0.3)& 6.0 (0.3)& 9.6 (0.0)\rule[-2mm]{0mm}{6mm}\\ \hline
   \multicolumn {4} {c} {Non-circular motion of G5.89-0.39}\\ \hline
   ($\upsilon^{\prime}_{\mathrm{x}}$, $\upsilon^{\prime}_{\mathrm{y}}$, $\upsilon^{\prime}_{\mathrm{z}}$)&-12.8 (1.8)&0.4 (1.4)&4.7 (0.4)\rule[-2mm]{0mm}{6mm}\\
   ($U$, $V$, $W$)& 5.4 (1.6) & -5.5 (2.4) & 11.3 (2.4)\rule[-2mm]{0mm}{6mm}\\ \hline
   \multicolumn {4} {l} {$^a$ Velocities are in the units of km s$^{-1}$.}\\
     \end{tabular}
\end{table}

 The peculiar motion of the system have also been shown in ($U$, $V$, $W$) frame at the same time. 
 Here, $U$ is toward the Galactic center from the source position, 
 $V$ is along the Galactic rotation at the local place, $W$ is perpendicular to the Galactic plane and toward the north Galactic pole. 
 Magnitude of the non-circular component $\sim$ 14 km s$^{-1}$ is enough within theoretically predicted range (see section 5.1 again). 
 
 \subsubsection{Anisotropic Expansion of UCH$\emissiontype{II}$ Region}
 Table \ref{tb:UCHIIregion} summarizes the parameters of the UCH$\emissiontype{II}$ region. 
 Upper seven rows present the electron density $n_{\mathrm{e}}$, inner / outer radius of the radio shell $R_{\mathrm{i}}$ / $R_{\mathrm{o}}$, 
 expansion velocity $\upsilon_{\mathrm{shell}}$, dynamical age $t_{\mathrm{dyn}}$, 
 thermal energy of ionized gas $E_{\mathrm{therm}}$, kinetic energy $E_{\mathrm{kin}}$ and total momentum of shell expansion $P_{\mathrm{shell}}$. 
 Lower three rows contain the parameters of surrounding dense envelope. 
 Envelope mass $M_{\mathrm{env}}$ and density $n_{\mathrm{env}}$ are obtained from SMA observation of submillimeter dust continuum \citep{Tang2009} and its size scale is roughly 2 $\times$ 10$^{4}$ au. 
 In especial, envelope temperature $T_{\mathrm{env}}$ have been derived from an excitation analysis of the CH$_{3}$CN ($J$ = 12 -- 11) line by \citet{Su2009}, 
 and hence, do not depend on a distance. They have found a temperature gradient in the range shown in the table. 
 
 Overall properties of the ionized region are not changed drastically except for the expansion velocity 
 which is calculated from angular expansion rate of 4 $\pm$ 1 mas yr$^{-1}$ in ACW98. 
 Even if ionized gas is in a free expansion state, its expansion velocity should be almost equal 
 to a sound speed of ionized gas ($\sim$ 13 km s$^{-1}$ at 10$^{4}$ K). 
 This suggests that only a half of total momentum can be thermally provided and the shell must require another source of momentum input. 
 If we attribute it to a spherically symmetric stellar wind discussed in section 5.2.1, required time is about 1000 yr. 
 Moreover if the momentum of photons from Feldt's star is added, 
 the total momentum of the shell could be provided during $t_{\mathrm{dyn}}$. 
 
 ACW98 have determined the source distance in comparison between angular expansion rate and the line 
 width of radio recombination lines under several assumptions about how the shell expansion contributes to the line width. 
 Now we have more accurate distance, and in turn, 
 we can estimate a residual contribution to the line width which offsets from spherical expansion.
 This value is $\sim$ 50 km s$^{-1}$ based on their analysis.
 This is twice as large as tangential expansion and nearly consistent with highly blue shifted OH masers 
 in front of the shell. 
 
 This highly anisotropic expansion could be explained by the idea proposed by \citet{Hunter2008}, 
 where an outflow from SMA1, which is the luminous dust core near Feldt's star, had been going across north to south and it was disrupted after the formation of the UCH$\emissiontype{II}$ region.
 If this is the case, ionization of intrinsically expanding gas within the outflow can give a plausible explanation 
 for high velocity ionized gas, and hence, observed anisotropy. 
 
 \subsubsection{Outflow Nature}
 The revised properties of the outflowing gas in G5.89-0.39 are listed in tables \ref{tb:Ooutflow} and \ref{tb:Ioutflow}. 
 Table \ref{tb:Ooutflow} shows the parameters of CO ($J$ = 1 -- 0) emission which trace the outer extended flow \citep{Watson2007}, 
 and table \ref{tb:Ioutflow} presents that of the inner high velocity flow which has been detected in CO ($J$ = 3 -- 2) emission \citep{Klaassen2006}. 
 Each table contains the dynamical age $t_{\mathrm{dyn}}$, outflow velocity (or velocity ranges) $\upsilon_{\mathrm{flow}}$, outflow mass $M$, 
 kinetic energy $E_{\mathrm{kin}}$, momentum $P$, outflow rate $\dot{M}$, 
 momentum rate $\dot{P}$ and mechanical luminosity $L_{\mathrm{mech}}$. 
 We note that an effect of inclination is not corrected in these tables, 
 although \citet{Klaassen2006} have originally assumed inclination of $45^{\circ}$. 
 
 The largest outflow in table \ref{tb:Ooutflow} is thought to be driven by Feldt's star and extends just east -- west direction (see figure 2 in \cite{Watson2007}). 
 Total mass measured from CO ($J$ = 1 -- 0) is $\sim$ 100 $M_{\odot}$ and greatly exceeds the source mass. 
 Such an extremely massive outflow is often observed toward a high mass star-forming region (e.g., \cite{Lopez2009}). 
 The origin of large mass is unsolved problem (e.g., \cite{Churchwell1997}). 
 
 A detailed driving mechanism of a protostellar outflow is still under discussion, however \citet{Machida2008} have shown that 
 high velocity jet and molecular outflow (hereafter called intrinsic outflows) are distinctly driven in their MHD simulation 
 as natural products of magnetized core collapse and mass accretion via rotating disk. 
 This is consistent very well with a case of nearby low mass star formation, 
 and seems to be applicable for high mass case as long as objects are formed by gravitational collapse and mass accretion. 
 The energy source of intrinsic outflows is gravitational energy which is transformed into outflow energy mediated by magnetic fields. 
 In this point of view, total outflow mass of intrinsic outflows cannot exceed that of central object at a driving point. 
 Clearly, much of the outflow mass arises from material entrained far from the driving source. 
 Quantitatively, observations of outflows from massive young protostellar objects show 
 that the entrained mass is on the order of 4\% of the core mass surrounding the central protostar (e.g., \cite{Beuther2002a}). 
 This relation suggests that the parent core for the forming cluster containing Feldt's star contained about $\sim$ 2500 $\MO$. 
 
 On the other hand, if we assume the ratio of the outflow to the accretion rate as $\sim$ 10 \% from theoretical works (e.g, \cite{Pelletier1992}), 
 total mass of intrinsic outflows for Feldt's star is $\sim$ 2.5 $\MO$ following the discussion in section 5.2.1. 
 This is comparable with the mass of the CO ($J$ = 3 -- 2) outflow which traces inner hot and high velocity outflow (see table \ref{tb:Ioutflow}). 
 If we consider a momentum transport from the CO ($J$ = 3 -- 2) outflow to the CO ($J$ = 1 -- 0) outflow, required time is 2.3 $\times$ 10$^{4}$ yr. 
 This is consistent with a formation time of Feldt's star estimated above and 
 clearly larger than the dynamical age of outflow which is obtained from an extent divided by a current velocity. 
 We note that this time scale should be a lower limit, 
 since the mass and momentum of the CO ($J$ = 3 -- 2) outflow is the sum of the east -- west flow from Feldt's star and north -- south flow from SMA1. 
 They have been unresloved in the single dish data in \citet{Klaassen2006}, 
 and actually, significant fraction of high velocity emission arises from the latter (see high resolution images in \cite{Hunter2008}). 
 
 The fact that a commonly used dynamical time underestimates actual outflow age 
 has already been pointed out in the case of low mass objects \citep{Parker1991} and we suggest that 
 it can occur in the case of too massive outflow in high mass star formation which includes large entrained mass. 
 It should be, consequently, careful that an observationally estimated outflow rate, 
 which often reaches 10$^{-2}$ \MO$\:$ yr$^{-1}$ for high mass star, is also overestimated by an order of magnitude. 
 This looks reasonable because \citet{Hosokawa2009} have predicted that large accretion 
 luminosity must prevent steady accretion in the case of such a too large accretion rate. 
 Detailed information about a termination of high velocity inner flow is quite useful to confirm these discussions. 
 This would be obtained from direct measurement of proper motions of outflow lobes with ALMA. 
 
\section{Conclusion}
 The distance of G5.89-0.39 is newly estimated to be 1.28$^{+0.09}_{-0.08}$ kpc from the annual parallax measurement with VERA. 
 This is 2/3 of the previously known value, but it is well reasonable 
 if we take into account the small galactic longitude of $\sim$ 5.89$^{\circ}$ and recent theoretical prediction about 
 non-circular motions of star-forming regions. 
 
 Rescaled physical parameters based on the new distance give us several in-depth natures of high mass star formation in G5.89-0.39 as follows. 
 
 (1) The ionizing star is rather later type ZAMS than previously believed type of O5. 
 Spectral type of O8.5 -- O8 means that the UCH$\emissiontype{II}$ region are excited by not so massive and standard O-type object. 
 Expected accretion rate is $\sim$ 10$^{-3}$ \MO$\:$ yr$^{-1}$ based on the extremely young age of the ionized shell and 
 detailed evolutional track of massive protostar under a high accretion rate. 
 Resultant formation time is about $\sim$ 2.5 $\times$ 10$^{4}$ yr in this case. 
 
 (2) Detected maser alignment at the O-star can be fitted by infalling Keplarian ring and its 
 inclination and position angle are also consistent with east-west orientation of the strong outflow. 
 It seems to be trace accretion disk (or its remnant) and suggest remarkable youth of the O-star 
 which is before complete evaporation of circumstellar structure.
 
 (3) Reconsideration of outflow nature suggests that the large portion of outflow mass should be entrained from massive envelope. 
 A commonly used dynamical time should significantly underestimate actual outflow age same as low mass cases \citep{Parker1991}. 
 This also causes an overestimate of outflow rate by an order of magnitude. 
 Direct observation of momentum transportation from intrinsic outflow to outer entrained flow is required to confirm this. 
 This may be able to be achieved with the proper motion measurement of outflow lobe with ALMA.

 We finally emphasize that G5.89-0.39 is one of the nearest target to 
 investigate individual high mass star formation and evolution of core scale cluster including an O-type object. 
 \\
 \\
 We would like to thank all the members of VERA project for their assistance in observations and data analyses. 
 We also thank the refree for helpful comments. 
 This work was financially supported by the Research Fellowships of the Japan Society for the Promotion of Science (JSPS). 

\bigskip






\end{document}